\documentclass{IEEEtran}
\usepackage{cite}
\usepackage{color}
\usepackage{amsmath,amssymb,amsfonts}
\usepackage{algorithm,algpseudocode}
\usepackage[caption=false]{subfig}
\usepackage{multirow}
\usepackage{tikz}
\usetikzlibrary{shapes,arrows,positioning, backgrounds,fit}
\usepackage{relsize}
\tikzset{fontscale/.style = {font=\relscale{#1}}
    }
\usepackage{graphicx}
\usepackage{textcomp}
\usepackage{bbm}
\usepackage{float}
\usepackage{setspace}
\usepackage{fancyhdr}
\usepackage{fancybox}
\usepackage[usenames,dvipsnames]{pstricks}
\usepackage{epsfig}
\usepackage{pst-grad} 
\usepackage{pst-plot} 
\usepackage{array}
\usepackage{varwidth}
\usepackage{xcolor}
\usepackage{tabularx}
\usepackage{diagbox}
\newtheorem{pb}{Problem}
\makeatletter
\newcommand{\multiline}[1]{%
  \begin{tabularx}{\dimexpr\linewidth-\ALG@thistlm}[t]{@{}X@{}}
    #1
  \end{tabularx}
}

\def\BibTeX{{\rm B\kern-.05em{\sc i\kern-.025em b}\kern-.08em
    T\kern-.1667em\lower.7ex\hbox{E}\kern-.125emX}}
\begin{document}
\title{Contrastive Blind Denoising Autoencoder for Real Time Denoising of Industrial IoT Sensor Data}
\author{Sa\'ul Langarica, \IEEEmembership{Student~Member, IEEE}, and Felipe N\'u\~nez, \IEEEmembership{Member, IEEE}
}


\maketitle
\begin{abstract}
In an industrial IoT setting, ensuring the quality of sensor data is a must when data-driven algorithms operate on the upper layers of the control system. Unfortunately, the common place in industrial facilities is to find sensor time series heavily corrupted by noise and outliers. In this work, a purely data-driven self-supervised learning-based approach based on a blind denoising autoencoder is proposed for real time denoising of industrial sensor data. The term \textit{blind} stresses that no prior knowledge about the noise is required for denoising, in contrast to typical denoising autoencoders where prior knowledge about the noise is required. Blind denoising is achieved by using a noise contrastive estimation (NCE) regularization on the latent space of the autoencoder, which not only helps to denoise but also induces a meaningful and smooth latent space. Experimental evaluation in both a simulated system and a real industrial process shows that the proposed technique outperforms classical denoising methods.
\end{abstract}


\section{Introduction}
The incorporation of Industrial Internet of Things (IIoT) technologies to modern industrial facilities enables the real-time acquisition of an enormous amount of process data, typically in the form of time-series, which represents an opportunity to improve performance by using data-driven algorithms for supervision, modeling and control \cite{lan:19,lan:21}.

Data-driven techniques, as statistical or machine learning algorithms, are capable of dealing with the multivariate and intricate nature of industrial processes; however, they rely on the consistency and integrity of the data to work properly \cite{nun:20}. This imposes an applicability limitation of these algorithms in real facilities, since IIoT sensor data is often highly corrupted with outliers and noise -caused by multiple factors as environmental disturbances, human interventions, and faulty sensors \cite{nun:20}- and there is a lack of effective real time cleaning techniques that can deal with these issues. As a result, the vast majority of successful data-driven case studies use offline preprocessed data, simulations or databases generated in a controlled environment;  applications in real industrial environments are scarce.

A typical practice for dealing with noisy process data is the use of smoothing filters \cite{industry_denoise}, like discrete low-pass filters, the Savitsky-Golay (SG) filter or exponential moving average filters (EMA). The main drawback of these techniques is their univariate nature, hence the redundancy and correlations among variables typically present in industrial processes are not exploited for denoising. Multivariate denoisers, which can exploit cross-correlation between signals, are a natural improvement to univariate filters. Approaches like Kalman or particle filters are the flagship techniques; however, they require the selection of a suitable model and a-priori estimation of parameters, as the covariance matrices in the Kalman filter.

A different approach is the use of transforms, like Wavelets \cite{wavelet} or Gabor \cite{gabor}, which exploit statistical properties of the noise so the signal can be thresholded in the transformed domain to preserve only the high-valued coefficients, and then, by applying the inverse transform, obtain a cleaner signal. A limitation of this technique is the difficulty of knowing a priori the best basis for representing the signals, and without knowledge on the noise nature, as is the case in real process data, is hard to determine where to threshold.

Learning-based denoising algorithms, like principal component analysis (PCA), Kernel PCA, or dictionary learning \cite{DL}, solve some problems of fixed transforms by learning a suitable representation of the data in a transformed space. These approaches are also multivariate in nature, hence exploit correlations between signals; nevertheless, they were designed for static data, e.g., images, and hence important information from temporal correlations are not exploited at all. 

Recently, denoising autoencoders (DAEs) \cite{DAE1,DAE2} have emerged as a learning-based denoising technique that is multivariate in nature and is capable of learning complex nonlinear structures and relationships between variables. This represents a great advantage over traditional learning-based techniques when dealing with highly nonlinear data. Originally, DAEs emerged for image denoising, but the use of recurrent neural networks has allowed their application in denoising dynamical data, such as audio and video \cite{RDAE1,RDAE2}. Unlike PCA, dictionary learning or fixed transforms techniques, DAEs are not blind in the sense that for learning to denoise a signal, the clean version of the signal (the target) has to be known beforehand. In addition, information about the characteristics of the noise affecting the signal is required to create realistic training examples. This is an important limitation for the use of DAEs in real-world applications where the clean version of the signal and the noise characteristics are usually unknown.

Motivated by the results in \cite{noise2noise} showing that, under some conditions, it is possible to recover clean observations by only looking at corrupted ones, and the recent advances in self-supervised learning \cite{NCE_review} that allow to produce high-quality and high-level representations of the data, in this work we propose a novel blind denoising autoencoder for real-time denoising of IIoT sensor data, called Contrastive Blind Denoiser Autoencoder (CBDAE). In the proposed CBDAE, noise contrastive estimation (NCE) \cite{NCE} is used as a regularization technique over the latent space of a recurrent autoencoder to achieve blind denoising of multivariate time series. The CBDAE preserves all the advantages of DAEs and eliminates the need of prior knowledge of noise characteristics and the clean version of the signals, i.e., it achieves blind denoising.

The contributions of this paper are three-fold. First, we introduce a novel technique, the CBDAE,  which uses NCE as a temporal regularization over the latent space enabling blind denoising, i.e., denoise without  prior knowledge of the noise. Second, we present a methodology for finding \textit{hard negative} examples, which is an active field of research in contrastive learning, to train the CBDAE more efficiently. Finally, we show that  NCE regularization induces smooth, meaningful and compact representations of input sequences in the latent space, which can be used for other downstream tasks such as fault detection or prediction in the latent space.

The rest of this manuscript is organized as follows. Preliminaries are given in Section II. In Section III the CBDAE is presented. In Section IV, the CBDAE is evaluated both in a simulated system and and using real IIoT sensor data from a real industrial Paste thickener. Finally, conclusions and future directions of research are given in Section VI.

\section{Preliminaries}
\label{sec.sec2}
\subsection{Notation and Basic Definitions}
In this work, $\mathbb{R}$ denotes the real numbers, $\mathbb{Z}_{\geq0}$ the nonnegative integers, $\mathbb{R}^n$ the Euclidean space of dimension $n$, and $\mathbb{R}^ {n\times m}$ the set of $n\times m$ matrices with real coefficients. For $a,b\in\mathbb{Z}_{\geq0}$ we use $[a;b]$ to denote their closed interval in $\mathbb{Z}$. For a vector ${v}\in\mathbb{R}^n$, $v^i$ denotes its $i$th component. For a matrix $\mathbf{A}\in\mathbb{R}^ {n\times m}$, $\mathbf{A}_i$ denotes its $i$th column and $\mathbf{A}^i$ its $i$th row. For an n-dimensional real-valued sequence $\alpha:\mathbb{Z}_{\geq0} \rightarrow \mathbb{R}^n $, $\alpha(t)$ denotes its $t$th element, and $\alpha_{[a;b]}$ denotes its restriction to the interval $[a;b]$, i.e., a sub-sequence. Similarly, $\alpha_{\sim r[a;b]}$ denotes a sub-sequence of length $r$ whose elements are randomly chosen from $\alpha_{[a;b]}$. For a sub-sequence $\alpha_{[a;b]}$, $\mathbf{M}(\alpha_{[a;b]})\in\mathbb{R}^ {n\times (b-a+1)}$ is a matrix whose $i$th column is equal to $\alpha(a+i-1)$, with $i\in[1;b-a+1]$. The same notation applies for an n-dimensional finite-length sequence $\alpha:[0;\bar{T}]\rightarrow \mathbb{R}^n$ with the understanding that for a sub-sequence $\alpha_{[a;b]}$, $[a;b]\subseteq [0;\bar{T}]$ must hold. Given an N-dimensional (sub-)sequence $\alpha$, we define its $T$-depth window as a matrix-valued sequence $\beta:\mathbb{Z}_{\geq0} \rightarrow \mathbb{R}^{N\times T} $, where $\mathbf{\beta}(t)= \mathbf{M}(\alpha_{[t-T+1;t]})$. Given a finite set of random samples $\mathcal{X}$, and a function $f:\mathcal{X}\rightarrow\mathbb{R}$, $\mathbb{E}_{\mathcal{X}}\left[f(\cdot)\right]$ denotes the estimated expectation of $f$ from $\mathcal{X}$.
\subsection{Autoencoders and Denoising Autoencoders}
Autoencoders (AEs) \cite{AE1} are unsupervised neural networks trained to reconstruct their inputs at the output layer, passing through an intermediate layer usually of lower dimension than the inputs.

Formally, given an $n$-dimensional sequence ${y}$, the AE maps an input vector ${y}(t)$ to a latent representation ${z} \in \mathbb{R}^{m}$, with $m < n$, using a function $f_{\theta_{E}}$, which in the simplest case is a linear layer with $\sigma$ as an arbitrary activation function, namely,
\begin{equation}
\mathbf{z} = f_{\theta_{E}} = \sigma(\mathbf{W}_{\theta_{E}}{y}(t) + {b}_{\theta_{E}}),
\end{equation}
where $\mathbf{W}_{\theta_{E}} \in \mathbb{R}^{m \times n}$ and ${b}_{\theta_{E}} \in \mathbb{R}^{m}$ are trainable parameters of the network. In more complex approaches, $f_{\theta_{E}}$ can be chosen to be any type of layer, like recurrent or convolutional layers, or even a stack of multiple layers. 

After encoding, the latent vector is mapped back to the input space by a second function $g_{\theta_{D}}$ known as the decoder: 
\begin{equation}
\mathbf{\hat{y}}(t) = g_{\theta_{D}} = \sigma(\mathbf{W}_{\theta_{D}}{z} + {b}_{\theta_{D}}),
\end{equation}
where $\mathbf{W}_{\theta_{D}} \in \mathbb{R}^{n \times m}$ and ${b}_{\theta_{D}} \in \mathbb{R}^{n}$. 

During training, the parameters of the AE are found by solving the following optimization problem:
\begin{equation}
\theta^{*} = \text{arg min}_{\theta} ||\mathbf{\hat{y}}(t) - \mathbf{y}(t)||^{2},
\end{equation}
where $\theta$ accounts for all the trainable parameters. Fig. \ref{fig:autoencoder} illustrates a simple AE network.

\def\layersep{2cm}
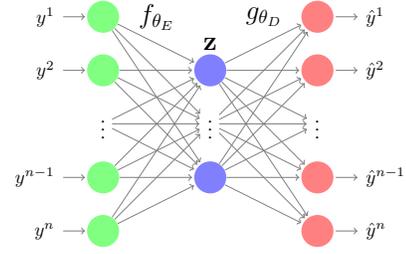
\begin{figure}
\begin{center}
\resizebox{0.3\textwidth}{!}{
\begin{tikzpicture}[shorten >=1pt,->,draw=black!50, node distance=\layersep]
    \tikzstyle{every pin edge}=[<-,shorten <=1pt]
    \tikzstyle{neuron}=[circle,fill=black!25,minimum size=17pt,inner sep=0pt]
    \tikzstyle{dots} = [rectangle]
    \tikzstyle{input neuron}=[neuron, fill=green!50];
    \tikzstyle{output neuron}=[neuron, fill=red!50];
    \tikzstyle{hidden neuron}=[neuron, fill=blue!50];
    \tikzstyle{annot} = [text width=4em, text centered]

    \node[input neuron, pin=left:$y^{1}$] (I-1) at (0,-1) {};
    \node[input neuron, pin=left:$y^{2}$] (I-2) at (0,-2) {};
    \node[dots] (I-3) at (0,-3) {$\vdots$};
	\node[input neuron, pin=left:$y^{n-1}$] (I-4) at (0, -4) {};
	\node[input neuron, pin=left:$y^{n}$] (I-5) at (0, -5) {};
	\node[annot] (A-2) at (1., -1) {\Large $f_{\theta_{E}}$};    
	
    \node[hidden neuron] (H-1) at (\layersep,-2) {};
    \node[dots] (H-2) at (\layersep,-3) {$\vdots$};
    \node[hidden neuron] (H-3) at (\layersep,-4) {};
	\node[annot] (A-1) at (\layersep, -1.5) {\Large $\mathbf{z}$};    
   
    \node[output neuron, pin={[pin edge={->}]right:$\hat{y}^{1}$}] (O-1) at (2*\layersep,-1) {};
    \node[output neuron, pin={[pin edge={->}]right:$\hat{y}^{2}$}] (O-2) at (2*\layersep,-2) {};
    \node[dots] (O-3) at (2*\layersep,-3) {$\vdots$};
    \node[output neuron, pin={[pin edge={->}]right:$\hat{y}^{n -1}$}] (O-4) at (2*\layersep,-4) {};
    \node[output neuron, pin={[pin edge={->}]right:$\hat{y}^{n}$}] (O-5) at (2*\layersep,-5) {};
	\node[annot] (A-3) at (3., -1) {\Large $g_{\theta_{D}}$};

    \foreach \source in {1,...,5}
        \foreach \dest in {1,...,3}
            \path (I-\source) edge (H-\dest);

	\foreach \source in {1,...,3}
        \foreach \dest in {1,...,5}
            \path (H-\source) edge (O-\dest);
       
\end{tikzpicture}}
\caption{Simple AE architecture where $f_{\theta_{E}}$ encodes the input data $y$ to a latent representation ${z}$ and then $g_{\theta_{D}}$ decodes ${z}$ to reconstruct the input as $\hat{y}$.}
\label{fig:autoencoder}
\end{center}
\end{figure}

When the input of the AE is corrupted with noise and the target output is clean, the resulting latent space is more robust, the AE learns richer features \cite{DAE4}, and learns how to denoise corrupted inputs. This is the working principle of the so called Denoising Autoencoders (DAEs), and has led to many denoising applications in static data, as images \cite{DAE5}, and, with the use of recurrent AEs \cite{RDAE1}, this technique has also been effectively applied to dynamic data \cite{RDAE2}. However, DAEs work under the assumption that the clean version of the input data is available (since is used as the target during training), as well as information about the noise. This limits the applicability of DAEs to IIoT sensor data where, in general, none of these requirements are fulfilled. 

\subsection{Noise contrastive estimation}
NCE \cite{NCE} is a learning method for fitting unnormalized models that has been adapted for different machine learning tasks, such as natural language processing (NLP) \cite{NCE_NLP}, time series feature extraction \cite{NCE_TS} and semi-supervised image and audio classification \cite{NCE_Image, NCE_Audio}. The basic idea of NCE is to estimate the parameters of the model by learning to discriminate between samples from the target distribution and samples from an arbitrary noise distribution, transforming complex density estimation tasks into probabilistic binary classification ones. NCE is based on ``learning by comparison'' and thus, falls under the family of contrastive learning methods \cite{NCE_review}.  

Mathematically, let ${y} \in \mathbb{R}^n$ be an input data vector belonging to one of $L$ possible classes. Denote as ${y}^{+}$ and ${y}^{-}$ positive and negative examples of ${y}$, respectively, meaning that ${y}^{+}$ and ${y}$ are of the same class, and ${y}^{-}$ belongs to a different class. Consider a generic nonlinear learnable transformation with parameters $\theta$, $f_{\theta}: \mathbb{R}^n \rightarrow \mathbb{R}^{m}$, which takes elements from the input data space and projects them to another space, i.e., an encoder. The NCE loss can be formulated as:
\begin{equation}
L_{NCE} = \mathbb{E}_{{y},{y}^{+},{y}^{-}} \left[-\log \left(\frac{e^{f_{\theta}({y})^T f_{\theta}({y}^{+})}}{e^{f_{\theta}({y})^T f_{\theta}({y}^{+})} + e^{f_{\theta}({y})^T f_{\theta}({y}^{-})}}\right) \right].
\end{equation}

The NCE loss has the property of maximizing a lower bound of the mutual information between ${y}$ and ${y}^{+}$ \cite{NCE_Audio}. However, as noted in \cite{NCE_multiple}, the NCE loss with only one negative example per update yields slow convergence and is prone to getting stuck in local optima. Therefore, the  NCE loss is modified in \cite{NCE_multiple} to use multiple negative examples, yielding a formulation known as the InfoNCE loss \cite{NCE_Audio}:
\begin{equation}
L_{NCE} = \mathbb{E}_{\mathcal{X}} \left[-\log \left(\frac{e^{f_{\theta}({y})^T f_{\theta}({y}^{+})}}{\sum_{y^k\in\mathcal{X}}e^{f_{\theta}({y})^T f_{\theta}({y}^{k})}}\right) \right],
\end{equation}
where $\mathcal{X}$ is a finite set of samples containing only one positive example: $y^+$. However, even if multiple negative examples are used, slow convergence might still be a problem, particularly when negative examples are far from positive ones, which produces a near-zero loss and very small gradients early in training. How to select the so-called \textit{hard negative} examples (those very close to the target but still farther than positive examples) is still an open problem in the literature.

Other proposals have been made to improve performance of the NCE loss, as using other similarity functions, e.g., cosine similarity, and introducing a temperature parameter \cite{NCE_Image}.


\section{Process data denoising using a Contrastive Blind Denoising Autoencoder}
\label{sec.sec4}

\begin{figure}
  \includegraphics[width=.95\linewidth]{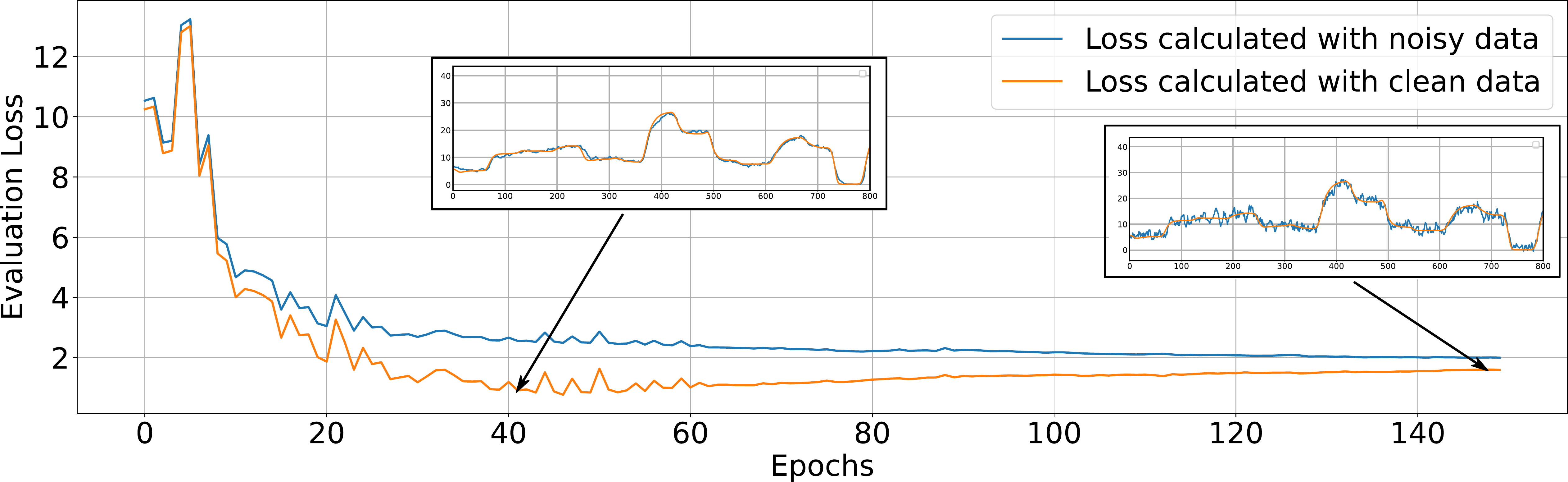}
   \centering
   \caption{Blind denoising with a recurrent AE with no regularization. The blue line represents the loss calculated using the output of the AE and the noisy signal, i.e., blind denoising, while the orange line represents the loss calculated using the output of the AE and the clean version of the signal (not available for training), which is the real objective to minimize, but in practice not possible to calculate. The network learns to denoise the signal at some point, but as training continues it begins to overfit, reproducing noise as well.}
  \label{fig:overfit}
\end{figure}

In this work we aim to transfer the blind denoising results for images presented in \cite{noise2noise} and \cite{blind_denoising_original} to time series, by using a recurrent AE. In particular, in \cite{noise2noise} it was shown that a neural network can be trained to denoise its inputs by only using corrupted versions as a target, i.e., blind denoise, as long as the noise is zero-mean. Nonetheless, it was also shown that when an AE with enough capacity is used, at some point it not only learns the dynamical characteristics of the system (which is fundamental for denoising), but also begins to overfit the data and reproduces noise as well. Figure \ref{fig:overfit} shows this effect when training a vanilla recurrent AE to denoise time series. 

To solve this issue, we propose to use NCE regularization to transfer temporal smoothness from the input time series to the latent space of the AE. NCE regularization forces the latent space vectors to encode the underlying shared information between different parts of the input signals, i.e., dynamical characteristics of the process, leaving out low-level information, as noise \cite{NCE_Audio}. Additionally, NCE regularization contributes to obtaining meaningful and compact representations of the input time series, which can be used for other downstream tasks, as it has been shown in the literature \cite{NCE_videos}.

\subsection{General setup}
Consider an underlying multi-input multi-output dynamical process, which is sampled periodically by an IIoT system with period $\tau$. From the IIoT system perspective, the process can be modeled as a discrete-time system given by
\begin{equation}
{x}(n + 1) = {F}_n({x}(n), {w}(n)).
\label{eqn.process}
\end{equation}

Where $n$ denotes the time step, ${x}(n)\in\mathbb{R}^W$, with $W$ the order of the system, ${x}(n)$ denotes the current value of the internal state, ${x}(n + 1)$ is the future value of the state, ${w}(n)$ denotes process inputs and ${F}$ is a smooth nonlinear mapping governing the dynamics of the system.
 
The measurement (observation) process is given by
\begin{equation}
{\tilde{y}}(n) = {C}_n({x}(n), {v}(n)),
\end{equation}
where ${\tilde{y}}(n)$ denotes the set of variables measured by the IIoT system, ${C}_n$ is the output mapping, and ${v}(n)$ represents measurement noise. For a noise-free condition, we denote the noise-free (unknown) measured variables at time step $n$ as:
\begin{equation}
{y}(n) = {C}_n({x}(n), 0).
\end{equation}

Under this setup, the measurement process generates a sequence ${\tilde{y}}:\mathbb{Z}_{\geq0} \rightarrow \mathbb{R}^N $, where, without loss of generality, $N\neq W$, therefore:
\begin{equation}
\mathbf{\tilde{y}}(n) = [\tilde{y}^{1}(n), \tilde{y}^{2}(n), \dots, \tilde{y}^{N}(n)]^{T}\in \mathbb{R}^{N},
\end{equation}
where each $\tilde{y}^{i}(n)$ represents a real-valued measurement of the $i$th variable at time step $n$. 

\subsection{Problem formulation}
Let $\mathbf{\tilde{Y}}$ be the $T$-depth window of ${\tilde{y}}$, i.e.,
\begin{equation}
\mathbf{\tilde{Y}}(n) = [{\tilde{y}}(n - T +1), {\tilde{y}}(n - T + 2), \dots,  {\tilde{y}}(n)] \in \mathbb{R}^{N \times T},
\label{eqn:window_eq}
\end{equation}
hence $\mathbf{\tilde{Y}}_k(n)\in \mathbb{R}^{N}$ is a column vector containing the $N$ measurements at time instant $n-T+k$, and $\mathbf{\tilde{Y}}^k(n)\in \mathbb{R}^{T}$ is a row vector containing the last $T$ samples of variable $k$.

The CBDAE uses $\mathbf{\tilde{Y}}(n)$ to generate an estimate of $y(n)$ at each time step $n$. Since the CBDAE is a dynamical system itself that processes data sequentially (due to the use of recurrent neural networks) to accomplish this task, two timescales should be introduced in the derivations. The  \textit{process time} refers to the time scale at which (\ref{eqn.process}) evolves and is indexed by $n$, while the \textit{CBDAE time} refers to the internal time of the CBDAE, which is indexed by $j$ for each fixed $n$. 

Since the objective of the CBDAE is to operate in real-time, the \textit{CBDAE timescale} has to be fast enough to process the data in $\mathbf{\tilde{Y}}(n)$ before another set of measurements, generated at \textit{process timescale}, becomes available. Hence, we have two dynamical systems at work, the slow system (the process) and the fast system (the CBDAE), which at each process time step gets triggered and iterates itself a number of steps depending on the size of $\mathbf{\tilde{Y}}(n)$ and the CBDAE architecture.

At this point, the problem of interest is formally introduced.

\begin{pb}
 Consider a dynamical process ruled by (\ref{eqn.process}) and an IIoT system generating a noisy measurement sequence $\tilde{y}$. At each time step $n$, generate an estimate $\hat{y}(n)$ of the noise-free measurement vector $y(n)$.
\end{pb}

Note that, unlike DAEs, in this case we do not have access to ${y}(n)$; therefore, denoising  must be done in a \textit{blind} manner.

\subsection{Blind Denoising}
The methodology for blind denoising will be divided in three fundamental steps, which are detailed in the following.

\subsubsection{Design of the Autoencoder network}
The following description is given for a fixed $n$, therefore we will drop this index for the sake of clarity. However, is important to remember the existence of the two timescales.

First, at process time $n$, the encoder generates a latent representation $h^L$ of $\mathbf{\tilde{Y}}$, which is a $Q_L$-dimensional finite-length sequence of length $T+1$ (an initial state plus one element for each vector of $\mathbf{\tilde{Y}}$), by feeding the network iteratively with the elements of $\mathbf{\tilde{Y}}$ and then recursively feeding back the internal state of the network, namely:

\begin{equation}
{h^L}(j) = f_{enc}(\mathbf{\tilde{Y}_j}, {h^L}(j - 1))
\label{eqn:encoder}
\end{equation}

where $f_{enc}$ represents a recurrent network, with $L$ layers and $Q_l$ neurons in each layer $l$, that in our case is composed by Gated Recurrent Units (GRU) \cite{gru}. 

 The operations at each GRU layer are given by:
\begin{equation}
\begin{aligned}
{z^l}(j) = \sigma(\mathbf{W_z}^l {p^l}(j) + \mathbf{U_z}^{h^l}(j) + {b_z}^l) \\
{r^l}(j) = \sigma(\mathbf{W_r}^{p^l}(j) + \mathbf{U_r}^{h^l}(j) + {b_r}^l) \\
{n^l}(j) = \tanh(\mathbf{W_n}^{p^l}(j) +{r^l}(j)\circ (\mathbf{U_n}^{h^l}(j) + {b_n}^l)) \\
{h^l}(j) = (1 - {z^l}(j)) \circ {n^l}(j) + {z^l}(j) \circ {h^l}(j),
\end{aligned}
\end{equation}
where ${h^l}(j) \in \mathbb{R}^{Q_{l}}$ is the hidden state of layer $l$ at CBDAE time $j$, ${r^l}$, ${z^l}$ and ${n^l}$ are $Q_l$-dimensional finite-length sequences representing the value of reset, update and new gates of layer $l$ respectively. $\sigma$ and $\tanh$ are the sigmoid and the hyperbolic tangent activation functions, $\circ$ is the Hadamard product and $p^l(j)$ is the input  of layer $l$ at CBDAE time $j$. For the first layer ($l=1$) the input, ${p^1}(j)$ is given by the input of the network $\mathbf{\tilde{Y}_j}(n)$, and for the subsequent layers the input is given by ${h^{l - 1}}(j)$, which is the current hidden state of the previous layer.

In this case, the latent representation of the input time series is selected to be ${h^{L}}(T) \in \mathbb{R}^{Q_{L}}$, which is the last hidden state of the last layer. This vector will be denoted as ${h}(T)$ from now on by dropping superscript $L$ for the sake of clarity.  

After the final latent representation is obtained, the decoder takes the final state of each encoder layer ${h}^i(T)$ as the initial state of each of its recurrent layers (${d}^i(0) = {h}^i(T)$) and begins to decode the sequence as
\begin{equation}
{d}^L(j) = f_{dec}({p_j}, {d}^L(j-1))
\end{equation}

where ${p_{j}} \in \mathbb{R}^{N}$ is the input for each decoding step of the network. This input could be either the delayed target $\mathbf{\tilde{Y}}_{j-1}$ or the previous network estimate $\mathbf{\hat{Y}_{j-1}}$. The network estimates are calculated by projecting ${d^{L}}(j)$ using a linear transformation $o_{dec}$ as
\begin{equation}
\mathbf{\hat{Y}_{j}} = o_{dec}({d^{L}}(j)) = \mathbf{W_{out}}{d^{L}}(j) + {b_{out}},
\label{eqn:decoder}
\end{equation}
where $ \mathbf{W_{out}} \in \mathbb{R}^{N \times Q_{L}}$ and $\mathbf{b_{out}} \in \mathbb{R}^N$ are trainable parameters. Then, the final output of the CBDAE corresponds to $\hat{y}(n)=\mathbf{\hat{Y}_{T}}$. Note that the decoder starts its processing once $\mathbf{h}(T)$ is available, which involves $T$ previous iterations of the encoder at CBDAE time. Therefore, careful should be taken when interpreting the local time of the decoder.

Two modifications are proposed to this vanilla version of the recurrent AE to enhance its denoising capabilities. First, it was found that \textit{scheduled sampling} \cite{scheduled} is beneficial for denoising. This technique consists in using at early training stages the target at $j - 1$ ($\mathbf{\tilde{Y}_{j - 1}}$) as the input of the decoder to predict the target at $j$, following (\ref{eqn:decoder}), and as training progresses, using network predictions at the decoder ($\mathbf{\hat{Y}_{j-1}}$) to predict the target at $j$. This gradual shift is controlled by a probability $p_d$, that in our case is increased linearly as
\begin{equation}
p_d = min(1, k_d + c_de),
\end{equation}
where $k_d$ and $c_d$ are parameters and $e$ is the training epoch. 

Shifting gradually from using the target as the input in each step to using the network's own predictions improves the stability of the training process, specially when the decoded sequence is long. Moreover, it forces ${h}(T)$ to retain all the information of the input sequence in a compressed manner, since the entire sequence has to be recovered from it, which in turn helps to retain only important information about the sequence, leaving out non-relevant information as noise. Although scheduled sampling contributes to denoising, it was experimentally found that by its own is not enough.

The second proposed modification, is the incorporation of an additional transformation that maps ${h}(T)$ to another space, where the contrastive loss is applied. This transformation is given by 
\begin{equation}
{z}(T) = g({h}(T)) = \mathbf{W}_{g2}\sigma(\mathbf{W}_{g1}({h}(T))) \in \mathbb{R}^{G},
\label{eqn:g_transform}
\end{equation}
where $\mathbf{W}_{g1} \in \mathbb{R}^{G_1 \times Q_{L}}$ and $\mathbf{W}_{g2} \in \mathbb{R}^{G \times G_1}$ are trainable parameters and $\sigma$ represents the Relu activation function. According to \cite{NCE_Image}, using this additional transformation is beneficial since the contrastive loss induces invariance of the representation between positive examples, hence applying it directly in the latent space could be detrimental for downstream tasks, where it is desirable that positive examples have similar but not equal representations.

\subsubsection{Selection of hard negative examples for NCE loss}
As mentioned, the incorrect selection of negative examples makes the network to exhibit sub-optimal performance and slow convergence \cite{mutual_information}. Therefore, is vital to develop an effective method for obtaining them, in the context of time series.

Although still an open problem, a typical approach for finding hard negatives is to randomly sample a batch of $B$ negative examples, calculate the score of each element and select the $K$ elements with higher loss values \cite{NCE_videos}.

Instead, we propose a hybrid scheme that takes advantage of the temporal consistency of the time series for selecting hard negatives. To this end, given a training database in the form of a finite length sequence $\mathbf{\tilde{Y}}$ of length $\bar{T}$, for a given index $k$, define a batch $\mathbf{B_s}$ as a sub-sequence $\mathbf{\tilde{Y}}_{[k-s+1;k]}$, i.e., a sub-sequence of data matrices, and, similarly, define a batch $\mathbf{B_r}$ as a sub-sequence of randomly chosen matrices $\mathbf{\tilde{Y}}_{\sim r[0;\bar{T}-1]}$. The batch for training is formed by concatenating both batches (sub-sequences) as
\begin{equation}
\mathbf{B} = \mathbf{B_s}\oplus\mathbf{B_r},
\end{equation}
where $\oplus$ denotes the concatenation operator. For each element of $\mathbf{B}$ perform: i) a forward pass of the encoder; ii) apply the transformation $g$ to the latent vector; and iii) concatenate the resulting projections to obtain the following matrix
\begin{equation}
\mathbf{Z} = [\mathbf{Z_s}\oplus \mathbf{Z_r}] \in \mathbb{R}^{G \times s+r}
\end{equation}
Since consecutive elements of $\mathbf{Z_s}$ are obtained by processing consecutive elements from $\mathbf{\tilde{Y}}$, the rationale is that their representations should be similar and very close in the latent space of the CBDAE. However, because of the temporal structure of the time series, it is desirable that $\mathbf{Z_s}_{j \pm 1}$ is closer to $\mathbf{Z_s}_{j}$ than $\mathbf{Z_s}_{j \pm 2}$, and $\mathbf{Z_s}_{j \pm 2}$ is closer than $\mathbf{Z_s}_{j \pm 3}$, and so on. Therefore, when we iterate through the batch, we can select $\mathbf{Z_s}_{j \pm 1}$ as a positive example of $\mathbf{Z_s}_{j}$ to increase the mutual information of consecutive representations, thus helping to transfer the dynamic characteristics of the underlying system to the latent space of the CBDAE. On the other hand, all the other representations $\mathbf{Z_s}_{j \neq \pm 1}$ are selected as negative examples to push them farther from $\mathbf{Z_s}_{j}$ than $\mathbf{Z_s}_{j \pm 1}$. These representations are inherently hard negatives because they will remain close due to the dynamical structure the space is learning. Similarly, $\mathbf{Z_r}$ representations are always used as negative examples.

Note that using only $\mathbf{B_s}$ with subsequent and highly correlated examples for training could damage learning since the batch is also used to train the autoencoding part \cite{dl_recomendations}. Also, as pointed out in \cite{mutual_information}, randomly selected samples could be beneficial in different stages of training when using contrastive losses, particularly in early stages. Therefore, we consider beneficial to use a combination of randomly selected and subsequent samples.

\subsubsection{Loss calculation}
After processing each batch and obtaining $\mathbf{Z}$, similar to \cite{NCE_Image}, the NCE loss is calculated using 
\begin{equation}
l(i, j, s) = -\log \left(\frac{\exp(sim(\mathbf{Z}_i, \mathbf{Z}_j))}{\sum_{k=1}^{B}\mathbbm{1}_{k \neq i, s}\exp(sim(\mathbf{Z}_i, \mathbf{Z}_k))} \right),
\label{eqn:nce_loss}
\end{equation}
where $sim$ stands for the cosine similarity, although, any similarity measure can be used, and $\mathbbm{1}_{k \neq i, s}$ is an indicator function evaluating to 1 if $k \neq i$ and $k \neq s$ and zero otherwise. To obtain the final NCE loss $L_{NCE}$, $l(i, j, s)$ is computed across all pairs $(j, j + 1)$ and $(j + 1, j)$ (since the NCE loss is not symmetric), where $j \in [1, s-1]$ because only $\mathbf{Z_s}$ elements are considered as positive pairs. Special attention should be given to index $s$, which is selected equal to $j - 1$ when the loss is calculated for the pair $(j, j + 1)$ and equal to $j + 2$ when the pair is $(j + 1, j)$. This is because when calculating the loss around $\mathbf{Z_j}$ with $\mathbf{Z_{j + 1}}$ as positive example, all other elements in $\mathbf{Z}$ are considered as negatives. However, we want representation $\mathbf{Z_{j - 1}}$ to remain as close to $\mathbf{Z_j}$ as $\mathbf{Z_{j + 1}}$, therefore $\mathbf{Z_{j - 1}}$ has to be excluded from the list of negatives. The same occurs when calculating the loss around $\mathbf{Z_{j + 1}}$, where representation $\mathbf{Z_{j + 2}}$ has to be excluded from the list of negatives.

Finally, the final loss of the batch consists in two elements, the autoencoding loss given by 
\begin{equation}
L_{AE} = \frac{1}{BNT}\sum_{k=1}^B\sum_{i=1}^{N}\sum_{j=1}^{T} |\mathbf{\hat{Y}_j^i}(k) - \mathbf{\tilde{Y}_j^i}(k)|,
\label{eqn:autoencoder_loss}
\end{equation}
and the NCE loss
\begin{equation}
L_{NCE} = \frac{1}{D} \sum_{k=1}^{D}\left( l(k, k +1, k-1) + l(k + 1, k, k+2) \right),
\label{eqn:nce_loss}
\end{equation}
where $D = 2s -1$. Finally, the batch loss is given by
\begin{equation}
L = L_{AE} + \beta L_{NCE},
\end{equation} 
where $\beta$ is a trade-off parameter between both terms. The complete training algorithm of the proposed CBDAE is presented in Algorithm \ref{alg:training}. 

\begin{algorithm}
\caption{Blind Denoising Autoencoder Training}
\label{alg:training}
\begin{algorithmic}[1]
\begin{small}
\State \textbf{Input}: Batch size $B$, number of training epochs $E$, training database size $\bar{T}$, number of measurements $N$, sequence length $T$, autoencoder transformations $f_{enc}$, $g$, $f_{dec}$ and $o_{dec}$, scheduled sampling parameters $p_d$, $k_d$ and $c_d$, batch ratio $r$ and loss parameter $\beta$.

\Procedure{CBDAE training}{}
\State Initialize $f_{enc}$, $g$, $f_{dec}$ and $o_{dec}$ weights randomly
\For{e in E}
\State $its = 0$
\While{$its \leq W$}
\State \multiline{Sample $s$ consecutive data matrices $\mathbf{\tilde{Y}}$ to form batch $\mathbf{B_s}$}
\State Randomly sample $r$ data matrices to form $\mathbf{B_r}$
\State $\mathbf{B} = \mathbf{B_s}\oplus\mathbf{B_r}$
\State Initialize $\mathbf{Z}$ and $\mathbf{\hat{B}}$ as empty matrices
\State $its = its + s$
\For {each $\mathbf{\tilde{Y}}$ in $\mathbf{B}$}
\State \multiline{Process $\mathbf{\tilde{Y}}$ iteratively with $f_{enc}$ as in (\ref{eqn:encoder}) to obtain $h(T)$}
\State \multiline{Use $g$ to obtain $z(T)$ from $h(T)$ as in (\ref{eqn:g_transform}) and save this representation in $\mathbf{Z}$}
\State \multiline{Initialize $\mathbf{\hat{Y}}$ as an empty matrix of autoencoder predictions}
\State Initialize state of the decoder $d(0) = h(T)$ 
\For{$j$ in $T$}
\State Select random number $\epsilon_d \in [0, 1]$
\State $SS$ = True If $\epsilon_d \leq p_d$ Else False
\If{$SS$ = True}
\State \multiline{$d(j) = f_{dec}(\mathbf{\hat{Y}_{j-1}}, d(j-1))$ (use past prediction)}
\Else
\State \multiline{$d(j) = f_{dec}(\mathbf{\tilde{Y}_{j-1}}, d(j-1))$ (use past target)}
\EndIf
\State $\mathbf{\hat{Y}_j} = o_{dec}(d(j))$
\EndFor
\EndFor
\State  \multiline{Calculate $L_{AE}$ with all the elements of $\mathbf{B}$ and $\mathbf{\hat{B}}$ as in  (\ref{eqn:autoencoder_loss})}
\State \multiline{Calculate $L_{NCE}$ with elements of $\mathbf{Z}$ as shown in (\ref{eqn:nce_loss})}
\State $L = L_{AE} + \beta L_{NCE}$
\State Update  $f_{enc}$, $g$, $f_{dec}$ and $o_{dec}$ to minimize $L$
\EndWhile
\State $p_d = min(1, k_d + c_de)$
\EndFor
\State Return $f_{enc}$, $g$, $f_{dec}$ and $o_{dec}$ that minimize $L$
\EndProcedure
\end{small}
\end{algorithmic}
\end{algorithm}

\section{Experimental Evaluation}
\subsection{Simulated example}
As a proof of concept, we use the CBDAE in a simulated industrial process, the well-known quadruple tank process \cite{quadruple_tank}, which is a nonlinear multi-input multi-output system that can switch between minimum phase and non-minimum phase behaviour. For all our experiments we used the non-minimum phase configuration to make the dynamics more challenging. Inputs are the voltages applied to the pumps, which vary in the range 0 to 1 volts. We assume that only the real value of the inputs (manipulated variables) are known, i.e., clean signals, which are normally available in any control system. The outputs are the four levels of the tanks that vary between 0 and 50 centimeters and are affected by noise. Both inputs and outputs are concatenated and given to the CBDAE as $\mathbf{\tilde{Y}}$. 

To gather data for training, the system is excited with multiple steps, then noise with different characteristics is added to the outputs and finally the CBDAE is trained to denoise the signals, following Algorithm \ref{alg:training}.

For all the experiments the following parameters were used: a sequence length of $T = 60$, batch size $B=64$, two hidden layers in the encoder and the decoder, and a hidden size of $80$ in each layer. The additional transformation dimension was selected as $dim(z(T)) = 20$ and the trade off parameter in the loss $\beta = 1.5$. Adam optimizer was used for optimization and Pytorch as the deep learning framework. Finally, the $RMSE$ between the original (clean) signal and the estimated clean signal is used as performance indicator for evaluation.

To evaluate blind denoising, the output signals are corrupted using a combination of white and impulsive noise, which is typically seen in real industrial processes. The noise power is varied from moderate ($\sigma = 0.5$) to very strong noise ($\sigma = 4$). To illustrate the advantages of the proposed CBDAE, we compare it against multiple classical data-driven and model-based baselines typically used in the industry, as well as against other BDAE networks trained with different variations of Algorithm \ref{alg:training}, this to highlight the positive effects of using the NCE loss.

As for the classical data-driven baselines, a Savitsky-Golay filter with a polynomial order of 2, an EMA filter with $\alpha = 0.33$ and typical mean and median filters were tested with different window sizes. For the classical model-based techniques, a Kalman Filter, a Particle Filter and an Extended Kalman filter, all using the real process matrices, were implemented as well. Finally, in the case of the different BDAE variations, we used a recurrent AE without any type of regularization (BDAE$_{NoReg}$) but trained as described in subsection III.C.1, a recurrent AE with $L_1$ regularization instead of NCE regularization (BDAE$_{L1}$) with $\beta = 1 \times 10^{-6}$ and a CBDAE without the transformation $g$ (CBDAE$_h$), meaning that in this case NCE regularization is applied directly on $h(T)$.

\begin{table*}
\caption{Comparison of denoising results in terms of the average RMSE [cm] for all the tank levels. For classical data-driven techniques $w_i$ represents the length of the window with which the best results were obtained.}
\begin{footnotesize}
	\begin{center}
    \begin{tabular}{| c | c | c | c | c | c | c | c | c |}
    \hline
    \backslashbox{Technique}{Noise Power}& $\sigma = 0.5$ & $\sigma = 1$ &  $\sigma = 1.5$ &  $\sigma = 2$ &  $\sigma = 2.5$ &  $\sigma = 3$ &  $\sigma = 3.5$ &  $\sigma = 4$\\ \hline \hline
    Original Noisy input & 2.165 & 2.406 & 2.572 & 2.850 & 3.121 & 3.515 & 3.929 & 4.277\\ \hline
    Mean F. ($w_5$) & 1.254 & 1.269 & 1.381 & 1.475 & 1.634 & 1.777 & 1.911 & 2.115\\ \hline
    Median F. ($w_5$)  & 0.803 & 0.917 & 1.117 & 1.337 & 1.583 & 1.785 & 2.019 & 2.173\\ \hline 
     Savitsky Golay F. ($w_{30}$) & 1.303 & 1.313 & 1.473 & 1.560 & 1.757 & 1.907 & 2.04 & 2.207\\ \hline
    EMA F.   & 1.244 & 1.264 & 1.375 & 1.469 & 1.630 & 1.773 & 1.913 & 2.089\\ \hline       
    Kalman F.  & 2.005 & 2.108 & 2.176 & 2.259 & 2.324 & 2.445 & 2.548 & 2.581\\ \hline
    Particle F. & 1.708 & 1.774 & 1.986 & 2.088 & 2.334 &  2.560 & 2.797 & 3.026 \\ \hline
    Extended Kalman F.  & 1.773  & 1.846 & 1.956 &  2.041 & 2.187 & 2.342 & 2.497 & 2.603 \\ \hline
    BDAE$_{NoReg}$  & 0.671 & 0.502 & 0.577 & 0.638 & 0.663 & 0.647 &   0.779 & 0.641\\ \hline
    BDAE$_{L1}$  & 0.668 & 0.572 & 0.58 & 0.587 & 0.588 &  0.668 & 0.566 & 0.620 \\ \hline
    CBDAE$_{h}$ & 0.392 & \textbf{0.313} & \textbf{0.326} & 0.445 & 0.521 & 0.569 & 0.741 &  0.602\\ \hline
    CBDAE  & \textbf{0.276} & 0.318 & 0.355 & \textbf{0.394} & \textbf{0.407} &  \textbf{0.518} & \textbf{0.489} & \textbf{0.542}\\ \hline
    \end{tabular}
      \label{table:results1}
     \end{center}
     \end{footnotesize}
\end{table*}

\begin{figure}
  \includegraphics[width=1\linewidth]{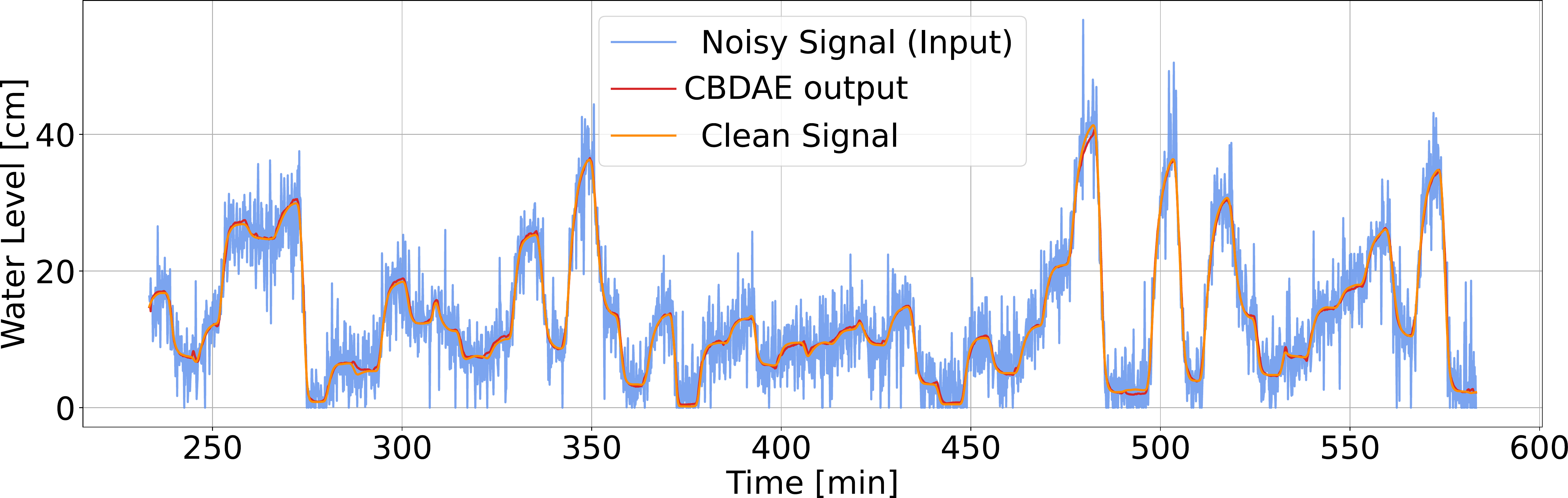}
   \centering
   \caption{CBDAE results when blind denoising one of the quadruple tank sensors corrupted with white and salt and pepper noise with $\sigma = 3$.}
  \label{fig:tanks_denoising_results}
\end{figure}

Table \ref{table:results1} shows the denoising results in terms of RMSE for the different methods. In the table, it can be seen that all BDAE methods clearly surpass both classical data-driven and model-based denoisers. In the case of data-driven denoisers, surprisingly the Savitsky Golay filter had the worst results. This is most likely due to the high frequency preservation property of this filter, which is specially harmful when the input signal is corrupted with ``salt and pepper'' noise. As for model-based denoisers, their poor performance may be due to the fact that these type of filters have major difficulties when faced with a non-minimum phase system \cite{noMinimum}. Regarding the different BDAE methods, it can be seen that the CBDAE$_h$ and the CBDAE, the two methods that use NCE regularization, achieve the best results in all cases. It is interesting to note that the CBDAE clearly outperforms the CBDAE$_h$ for high values of $\sigma$, which is probably due to the fact that the latent vectors of the CBDAE can make better use of the dynamical information of the input signals, thanks to transformation $g$, which apparently is increasingly important for high levels of noise. Fig. \ref{fig:tanks_denoising_results} shows the denoising results of the CBDAE when applied to denoise one of the quadruple tank signals.

\begin{figure*}
  \includegraphics[width=1\linewidth]{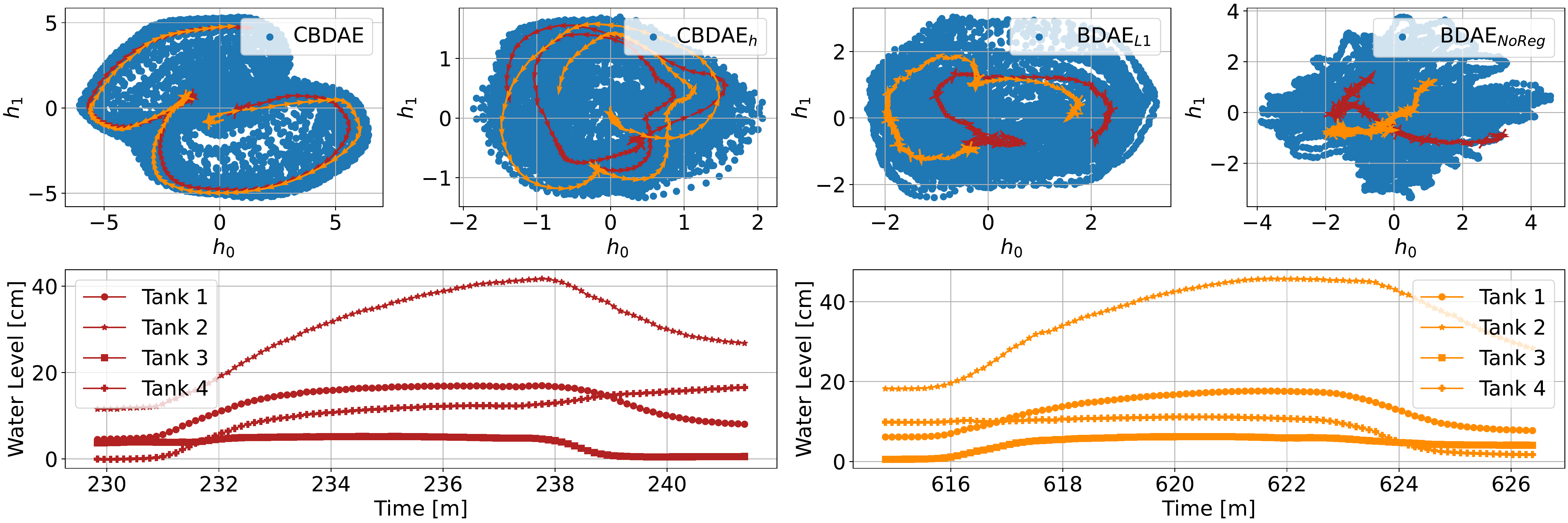}
   \centering
   \caption{Two-dimensional latent space trajectories obtained with PCA for the different BDAE networks when the process follows two similar trajectories for $\sigma = 3$. In the upper graphs, the red line represents the trajectory when the tank levels follow the lower left-hand graph trajectories and the orange line in the upper graphs represents the same for the lower right-hand graph.}
  \label{fig:latent_space_denoising}
\end{figure*}

Fig. \ref{fig:latent_space_denoising} shows the latent space trajectories obtained for the different BDAE networks when the input trajectories are very similar, after using PCA as dimensional reduction technique to project the latent space to two dimensions. It can be seen in the Figure that the CBDAE latent space is much more ordered, smooth and predictable than the latent space of the other BDAE networks. It is also interesting to note that the principal components of the CBDAE latent space present two lobes that match with the first-order information of the input signals. When the tanks are being filled, the latent space trajectory is on the right-side lobe, and when the tanks are being emptied, the latent space trajectory is on the left-side lobe. This dynamically smooth and ordered structure acquired by the latent space, as a result of using NCE regularization, could be further exploited for other downstream tasks as prediction and fault detection.

\subsection{Application to an industrial paste thickener}
As a second experiment, the CBDAE was applied to denoise the signals of a real industrial paste thickener. This process is considerable more challenging than the quadruple tank process, since the thickener is subject to many unmeasured disturbances and we only have access to a reduced number of measurements. Is worth mentioning that since in this experiment we are working with real process data, we do not have access to the clean version of the signals and therefore, we cannot evaluate the results of the CBDAE quantitatively, we will only inspect them visually. A detailed description of the IIoT infrastructure used to generate the data of the industrial paste thickener can be found in \cite{nun:20}.

\subsubsection{Thickening process description}
\begin{figure}
  \includegraphics[width=.95\linewidth]{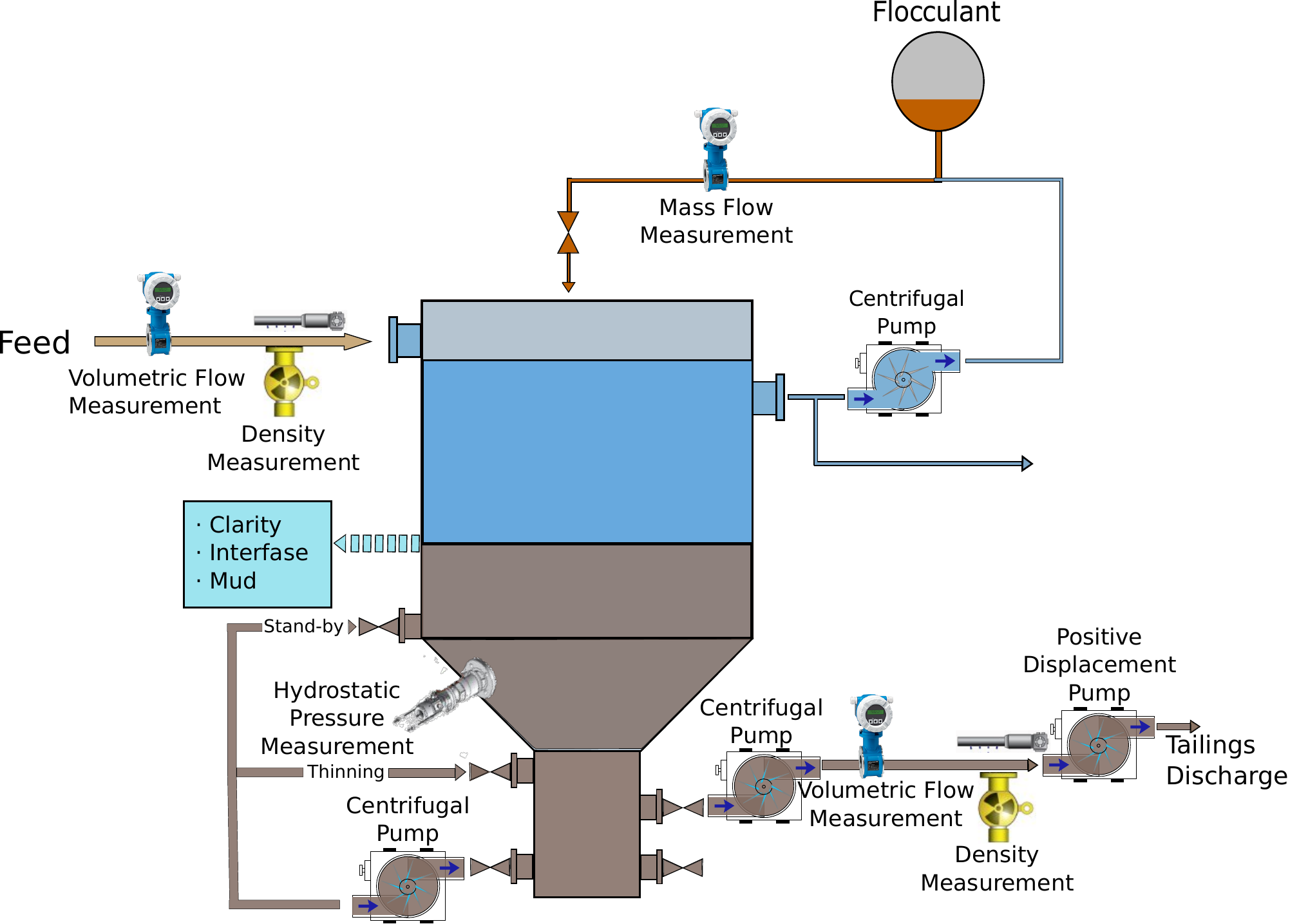}
   \centering
   \caption{Process and instrumentation diagram of the thickener used. Time series are available for the following variables: feeding and discharge rates, flocculant addition rate, feeding and discharge density, internal states of the thickener (mud, interface and clarity levels).}
  \label{fig:espesador}
\end{figure}

Thickening is the primary method in mineral processing for producing high density tailings slurries. Thickening generally involves a large tank (see Fig. \ref{fig:espesador}) with a slow turning raking system. Typically, the tailings slurry is added to the tank after the ore extraction process, along with a sedimentation-promoting polymer known as flocculant, which increases the sedimentation rate to produce thickened material discharged as underflow \cite{nun:20}.

Due to its complexity and highly non-linear dynamics, deriving a first-principles-based mathematical model is very challenging. Therefore, an appealing approach is to use data-driven modeling techniques. However, sensors in charge of providing data are exposed to strong disturbances and noise and an effective online preprocessing technique is needed. Hence, the thickener is an interesting real process to test the proposed CBDAE.

The CBDAE was trained with 12 months of real operational data from the industrial thickener, and following the guidelines of \cite{nun:20}, we selected 8 key variables that have been used to model this process with data-driven techniques. Namely, the flocculant flow, the output flow (manipulated variables), the input solids concentration, input flow (measured disturbances), the bed level, the rake torque, the hydrostatic pressure and output solids concentration (process outputs). 

\begin{figure}
  \includegraphics[width=1\linewidth]{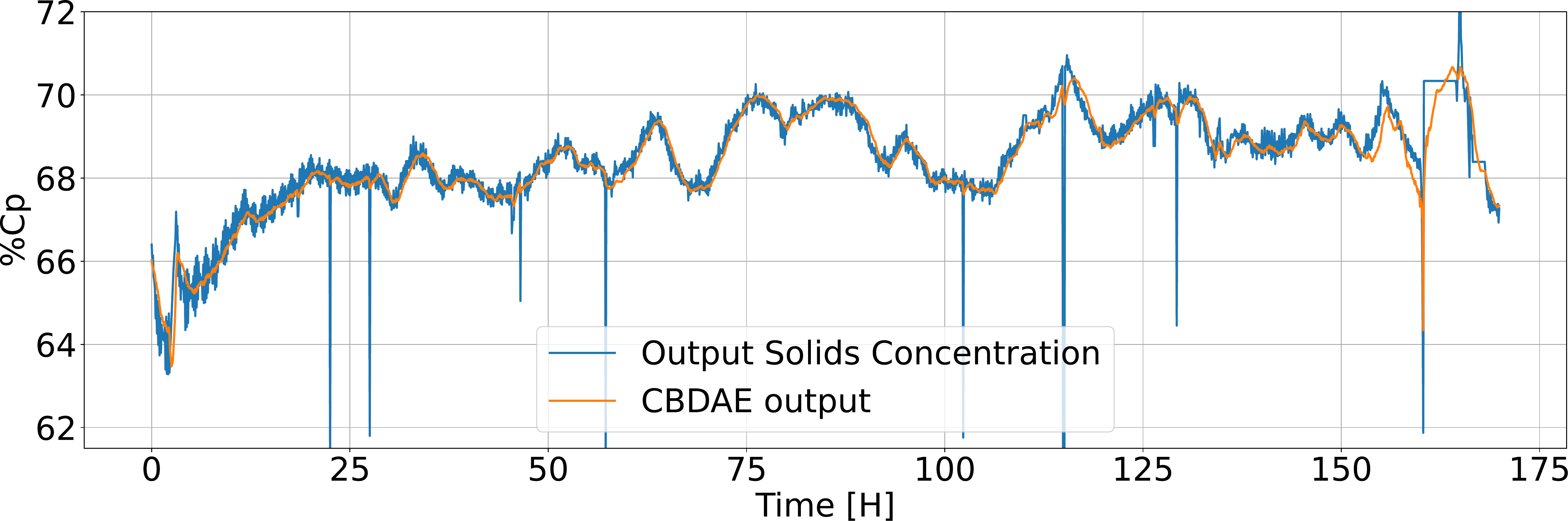}
   \centering
   \caption{CBDAE results when blind denoising the output solids concentration, one of the key variables of the thickener for control purposes.}
  \label{fig:thickener_denoising_results}
\end{figure}

\begin{figure}
  \includegraphics[width=1\linewidth]{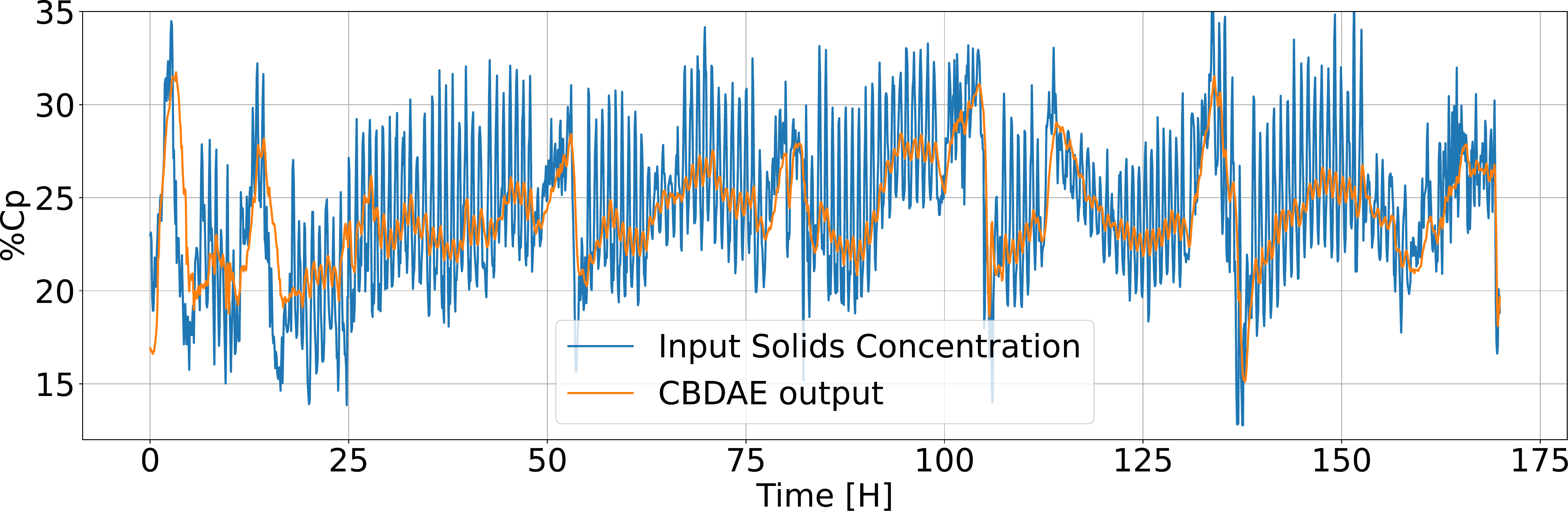}
   \centering
   \caption{CBDAE results when blind denoising the input solids concentration, a measured disturbance with one of the most corrupted sensors.}
  \label{fig:thickener_denoising_results2}
\end{figure}

\begin{figure}
  \includegraphics[width=1\linewidth]{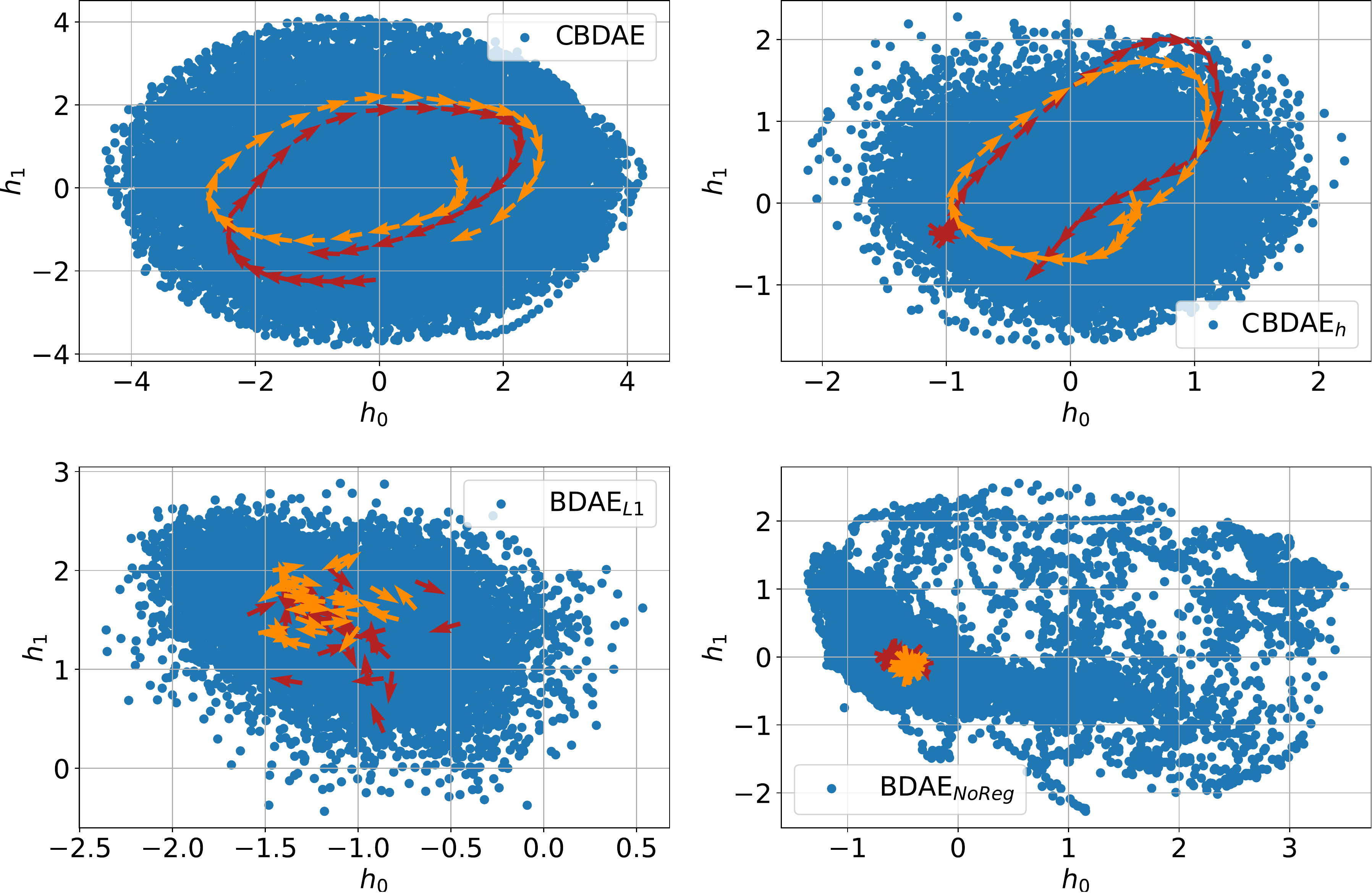}
   \centering
   \caption{Two-dimensional latent space trajectories obtained with PCA for the different BDAE networks when the thickening process follows two similar trajectories.}
  \label{fig:latent_space_denoising_thickener}
\end{figure}

Figs. \ref{fig:thickener_denoising_results} and \ref{fig:thickener_denoising_results2} show the denoising results of the CBDAE for the output solids concentration and the input solids concentration, which are two of the most important signals for control purposes. It can be seen in Fig. \ref{fig:thickener_denoising_results} that the CBDAE successfully learns to ignore the spikes in the signal, which clearly do not belong to the dynamics of the system. Similarly, in Fig. \ref{fig:thickener_denoising_results2}, it can be seen that the CBDAE is capable of filtering the strong noise affecting the input solids concentration. It is also interesting to note that the resulting delay is minimum, which could be critically important for control applications built in top. Analogous to Fig. \ref{fig:latent_space_denoising}, Fig. \ref{fig:latent_space_denoising_thickener} shows the trajectories in the latent space for two similar trajectories in the input space. Note that both the CBDAE and the CBDAE$_h$ show smooth and similar trajectories, unlike the networks that do not use NCE regularization, which show an erratic and totally unpredictable behavior.


%

\section{Conclusion}
In this work, the novel Contrastive Blind Denoising Autoencoder (CBDAE) is introduced as an alternative to classical methods for denoising corrupted time series obtained from an underlying dynamical system. The CBDAE is based on the use of NCE regularization in the loss function and the selection of hard negatives exploiting the temporal consistency of the corrupted time series. 

Experimental results show that the CBDAE outperforms classical data-driven and model-based denoising techniques typically used in the industry, as well as state of the art neural networks approaches. The use of the NCE regularization enables the CBDAE to capture high-level information, i.e., dynamical information, in the latent space leaving out irrelevant information, as noise. Additionally, it was shown that NCE regularization induces a smooth and meaningful structure in the latent space, which can be eventually used for other downstream tasks.

Future work includes testing the CBDAE approach for signal reconstruction, prediction and fault detection in the latent space of the network. 



\label{sec.con}

\bibliographystyle{IEEEtran}
\bibliography{IEEEexample}
\end{document}